\documentclass[12pt]{article}
\title{\bf Heavy-light mesons spectrum from the nonperturbative QCD 
in the einbein field formalism}
\author{
Yu.S.Kalashnikova\thanks{e-mail: yulia@heron.itep.ru}, A.V.Nefediev
\thanks{e-mail: nefediev@heron.itep.ru}}
\date{\it Institute of Theoretical and Experimental Physics,\\
B.Cheremushkinskaya 25, 117218, Moscow, Russia}
\newcommand{\be}{\begin{equation}}
\newcommand{\ee}{\end{equation}}
\newcommand{\ds}{\displaystyle}
\newcommand{\low}[1]{\raisebox{-1mm}{$#1$}}

\begin{document}
\maketitle

\begin{abstract}
The spectrum of $B$ and $D$ mesons (including the low lying orbitally and 
radially excited 
states) is calculated using the quark-antiquark Hamiltonian derived from QCD
in the einbein field formalism. Spin-spin and spin-orbit terms
due to the confinement and OGE interactions are taken into account as
perturbations. Results for the masses and splittings are confronted to 
the experimental and recent lattice data and are demonstrated to be in 
a reasonable agreement with both. We find that the orbital excitations
with $l=2$ and $l=3$ for $D$ meson lie approximately in the same 
region as its first radial excitation that might solve the mystery of the 
extremely narrow $D(2637)$ state recently claimed by DELPHI Collaboration.
\end{abstract}

\section{Introduction}

In spite of a rather long history of discussion and many theoretical 
attacks, an interest to the properties of heavy-light mesons is still
very high. Indeed, heavy-light systems incorporate both properties 
of the light quarks which are extremely important for the physics of
Chiral Symmetry Breaking (CSB), one of the most challenging phenomenon of 
the nonperturbative QCD, and heavy degrees of freedom which allow 
application of such profound methods of investigation as Heavy Quark 
Effective Theory (HQET) or Operator Product Expansion (OPE). 
The two main sources of information, experiment and lattice simulations, 
deliver new data on the heavy-light mesonic spectra, including orbital and radial 
excitations of the $q\bar q$ pair, which strongly need theoretical
identification and description. In the present paper we calculate the
masses of several low lying $D$ and $B$ mesonic states 
using the method developed in \cite{DKS,Green} based on the 
Hamiltonian approach to the bound states problem in QCD. Starting with the 
quarks with current masses we intensively use the so-called einbein fields
formalism \cite{einbein,yuaein,we-ein} which not only considerably simplifies 
calculations, but allows to justify the treatment of the spin-spin and 
spin-orbit interactions as perturbations, naturally explaining the 
appearance of the \lq\lq constituent" quark mass. Results for the spectra
and splittings are compared with the experimental and lattice data and a 
good agreement is found. 

We demonstrate that in our model the proper account of the QCD string 
dynamics gives extra negative contribution into the energy of the orbital
excitations and makes the orbitally excited state ($l=2$, $n=0$ and $l=3$, $n=0$) 
share the region of masses with the first radial excitation ($l=0$, $n=1$).
The latter observation may resolve the problem of the 
resonance $D(2637)$ recently claimed by DELPHI Collaboration \cite{DELPHI} 
and which is 
under intensive discussion at present (see $e.g.$ \cite{discussion}). 
Indeed, in spite of the fact that in our model both above mentioned states 
lie somewhat higher than
the experimentally observed resonance, the latter can be identified with the
orbital excitation of charm (with $l=2$ or $l=3$) rather then with the first radial 
excitation ($n=1$) solving in such a way the   
problem of an extremely small width of the resonance
inconsistent with any estimates of the $2S$-state decays \cite{discussion}.

\section{Einbein fields formalism and the mesonic Hamiltonian}

The einbein fields formalism which has a long history in 
literature \cite{einbein} was introduced as a method allowing to treat the
kinematics of relativistic particles, but then it was generalized for the case of
spinning particles and strings \cite{generalized,spin}. One of its applications
is the possibility to develop the canonical quantization
procedure $a\;la$ Dirac \cite{Dirac,we-ein}. In its simplest form this formalism can be applied to the
point-like scalar relativistic particle described by the action
\be
S=\int_{t_i}^{t_f}Ldt\quad L=-m\sqrt{\dot{x}^2}\quad
\dot{x}_{\mu}=\frac{\partial x_{\mu}}{\partial t}
\label{1}
\ee
so that the modified form of (\ref{1}) looks like
\be
L=-\frac{\mu\dot{x}^2}{2}-\frac{m^2}{2\mu},
\label{2}
\ee
where the einbein field $\mu$ is introduced. Dynamics defined by the equations
of motion for Lagrangians (\ref{1}) and (\ref{2}) is the same if the einbein 
field is treated as an independent degree of freedom and the corresponding
constraints are introduced and properly treated. For a short review of the einbein
fields formalism see $e.g.$ \cite{we-ein} and references therein. In the mean time there
exists 
another approach to the einbein fields \cite{yuaein} which allows to neglect 
their dependence on
the proper time $t$ and thus to treat them as variational parameters to be got
rid of in the spectrum rather than in the Hamiltonian. Such an approach was 
used to a success in a set of papers (see $e.g.$ \cite{yuaein,MNS}) and proved to be rather
accurate reproducing the relativistic spectra with the error within several per
cent. In what follows we accept the above accuracy and treat einbeins 
variationally. 

As it was shown in \cite{Green,DKS} within the Vacuum Correlators 
Method (VCM) \cite{VCM}, writing the gauge invariant Euclidean Green's function of 
the $q\bar q$ meson 
\be
G_{q\bar q}=\langle\Psi_{q\bar q}^+(\bar{x},\bar{y}|A)^+\Psi_{q\bar
q}^+(\bar{x},\bar{y}|A)\rangle_A
\quad\Psi_{q\bar q}(x_1,x_2|A)=\bar{\Psi}_{\bar q}(x_1)\Phi(x_1,x_2)\Psi_q(x_2),
\ee
with $\Phi(x_1,x_2)$ being the standard path-ordered parallel transporter 
operator
$$
\Phi(x_1,x_2)=P\exp{\left(ig\int_{x_2}^{x_1}dz_{\mu}A_{\mu}\right)}
$$
and employing Feynman--Schwinger representation for the single particle Green's
functions and the 
area law for the isolated Wilson loop bounded by the quark and antiquark
trajectories, one can arrive at the following result \cite{Green}
\be
G_{q\bar q}=\int D\mu_1(t_1)D\mu_2(t_2)D\vec{x}_1D\vec{x}_2 e^{-K_1-K_2}
Tr\left[\vphantom{\int_0^T\frac{d}{d\mu}}\Gamma^{(f)}(m_1-\hat D)
\Gamma^{(i)}(m_2-\hat{\bar D})\times\right.
\label{3}
\ee
$$
\left.
P_{\sigma}\exp{\left(\int^T_0
\frac{dt_1}{2\mu_1(t_1)}\sigma^{(1)}_{\mu\nu}\frac{\delta}{i\delta
s_{\mu\nu} (x_1(t_1))}\right)}\exp{\left(-\int^T_0\frac{d t_2}{2\mu_2(t_2)}
\sigma^{(2)}_{\mu\nu}\frac{\delta}{i\delta s_{\mu\nu}(x_2(t_2))}\right)}
\exp{\left(-\sigma S\right)}\right],
$$
where $\mu_1$ and $\mu_2$ are the einbein fields, $K_i$ being the 
kinetic energies of the quarks 
\be
K_i=\int_0^Tdt_i\left(\frac{m_i^2}{2\mu_i}+\frac{\mu_i}{2}+
\frac{\mu_i\dot{\vec{x}}_i^2}{2}\right),\quad i=1,2,
\label{4}
\ee
$S$ stands for the minimal area swept by the quark-antiquark trajectories,
$\sigma_{\mu\nu}=\frac{1}{4i}(\gamma_{\mu}\gamma_{\nu}-
\gamma_{\nu}\gamma_{\mu})$ and $\delta/\delta s_{\mu\nu}$ denotes the 
derivative with respect to the element of the area $S$. The form of the 
kinetic energy (\ref{4}) immediately follows from the equation (\ref{2}) if the
laboratory gauge $t=x_0$ is fixed in the proper time
reparametrization group. 
Equation (\ref{3}) allows to find the Hamiltonian of the
quark-antiquark pair connected with the relativistic string.
If one neglects spin-dependent terms in (\ref{3}), then the corresponding 
Hamiltonian coincides by the one for the Nambu-Goto string with scalar 
quarks at the ends which follows from the Lagrangian
\be
L(t)=-m_1\sqrt{\dot{x}_1^2}-m_2\sqrt{\dot{x}_2^2}
-\sigma\int_0^1d\beta\sqrt{(\dot{w}w')^2-\dot{w}^2w'^2}\quad x_{10}=x_{20}=t,
\label{5}
\ee
where the minimal string is parametrized by the profile function $w(t,\beta)$,
the dot and the prime denote its derivatives with respect to the proper
time $t$ and the string coordinate $\beta$ respectively. Thus for the
centre-of-mass frame one arrives at \cite{DKS}
$$
H=\sum_{i=1}^2\left(\frac{p_r^2+m_i^2}
{2\mu_i}+\frac{\mu_i}{2}\right)+\int^1_0d\beta\left(\frac{\sigma^2r^2}{2\nu}+
\frac{\nu}{2}\right)+\frac{\vec{L}^2}{2r^2[\mu_1{(1-\zeta)}^2+\mu_2{\zeta}^2+
\int^1_0d\beta\nu{(\beta-\zeta)}^2]},
$$
\be
\zeta=\frac{\mu_1+\int^1_0d\beta\nu\beta}{\mu_1+\mu_2+\int^1_0d\beta\nu}.
\label{6}
\ee

Note that an extra einbein $\nu$ following equation (\ref{6}) has the
meaning of the string energy density. Extrema in all three einbeins
($\mu_1$, $\mu_2$ and $\nu$) should be taken in (\ref{6}). For the case of zero
angular momentum $L=0$ the last term in (\ref{6}) vanishes and the extremal
value for $\nu$ is easily found to be
\be
\nu_0=\sigma r,
\label{7}
\ee
so that the resulting interaction gives the linearly rising potential $\sigma
r$. Expansion of the last term in Hamiltonian (\ref{6}) in powers of 
$\sqrt{\sigma}/\mu$ with the substitution of the extremal value $\nu_0$ yields
\be
H=H_0+V_{string}
\label{8}
\ee
\be
H_0=\sum_{i=1}^2\left(\frac{\vec{p}^2+m_i^2}{2\mu_i}+\frac{\mu_i}{2}\right)+\sigma
r-\frac43\frac{\alpha_s}{r}-C_0,
\label{9}
\ee
\be
V_{string}\approx -\frac{\sigma
(\mu_1^2+\mu_2^2-\mu_1\mu_2)}{6\mu_1^2\mu_2^2}\frac{\vec{L}^2}{r}
\label{10}
\ee
where we have supplied $H_0$ with the constant term $C_0$ and the 
colour Coulomb interaction which comes
from the perturbative gluon exchange omitted in (\ref{3}) for simplicity.
Equation (\ref{10}) gives the
first spin-independent correction called the string correction \cite{MP,DKS} which is known
to reduce the inverse Regge trajectory slope $dM_L^2/dL=4\sigma$ specific for the
heavy-light limit of the Hamiltonian $H_0$, and, when taken at the account 
at full scale, to lead to the correct string slope $\pi\sigma$ \cite{DKS,MNS}.

In what follows we shall consider $H_0$ as the zeroth approximation for
calculation of the mesonic spectrum. Once the angular momentum $\vec{L}$, the total 
spin of the quark-antiquark pair $\vec{S}$ and the total momentum
$\vec{J}=\vec{L}+\vec{S}$ are separately conserved with the Hamiltonian $H_0$ 
then one can
specify the mesonic states as terms $n^{2S+1}L_J$, where $n$ is the radial
excitation number. 

Let us turn back to the equation (\ref{3}) and to identify the spin-dependent
correction to the zeroth order Hamiltonian $H_0$. Following \cite{Green} one finds
$$
V_{sd}=\frac{8\pi\kappa}{3\mu_1\mu_2}(\vec{S}_1\vec{S}_2)
\left|\psi(0)\right|^2-\frac{\sigma}{2r}\left(\frac{\vec{S}_1\vec{L}}{\mu_1^2}+
\frac{\vec{S}_2\vec{L}}{\mu_2^2}\right)
+\frac{\kappa}{r^3}\left(\frac{1}{2\mu_1}+\frac{1}{\mu_2}\right)
\frac{\vec{S}_1\vec{L}}{\mu_1}
+\frac{\kappa}{r^3}\left(\frac{1}{2\mu_2}+\frac{1}{\mu_1}\right)
\frac{\vec{S}_2\vec{L}}{\mu_2}
$$
\be
+\frac{\kappa}{\mu_1\mu_2r^3}\left(3(\vec{S}_1\vec{n})
(\vec{S}_2\vec{n})-(\vec{S}_1\vec{S}_2)\right)+\frac{\kappa^2}{2\pi\mu^2r^3}
\left(\vec{S}\vec{L}\right)(2-{\rm ln}(\mu r)-\gamma_E),\quad \gamma_E=0.57
\label{11}
\ee
where $\kappa=\frac43\alpha_s$ and we have added the term of order $\alpha_s^2$
which comes from one-loop calculations and is extensively discussed in
literature \cite{alpha2,gluelump}.
Note that up to the last term potential (\ref{11}) coincides in form with the 
Eichten--Feinberg--Gromes results \cite{EFG}, but contains $\mu$'s in the denominators
instead of masses. In the meantime as clearly seen from (\ref{9})
$\mu_i\sim\sqrt{\langle\vec{p}^2\rangle+m_i^2}>m_i$ or even $\mu_i\gg m_i$ for
light quarks\footnote{Let us remind here that these are current quark masses to
be denoted $m_i$ and to enter equations (\ref{1})-(\ref{10})} (see also Table 1). Einbeins $\mu$'s can
be viewed as dynamical or \lq\lq constituent" quark masses that makes it
possible to consider (\ref{11}) as a correction and to justify the expansion made
in (\ref{10}) (see {\it e.g.} \cite{Lisbon} for more detailed discussion of this issue).

\section{Numerical solution}
\subsection{Spectrum of the Hamiltonian $H_0$}

As mentioned in the introduction einbein fields can considerably simplify 
the relativistic dynamics. Indeed with the einbeins $\mu$'s introduced
the kinetic part of the Hamiltonian $H_0$ from (\ref{9}) has a nonrelativistic
form, so that the corresponding Schr{\" o}dinger equation can be written in the
reduced dimensionless form (see $e.g.$ \cite{yuaDB,Lisbon})
\be
\left(-\frac{d^2}{d\vec{x}^2}+|\vec{x}|-\frac{\lambda}{|\vec{x}|}\right)
\chi_{\lambda}=
a(\lambda)\chi_{\lambda}\quad
\lambda=\kappa\left(\frac{2\mu}{\sqrt{\sigma}}\right)^{2/3}
\label{12}
\ee

On finding the solutions of
equation (\ref{12}) for $\chi_{\lambda}$ and $a(\lambda)$ as functions of the reduced Coulomb
interaction strength $\lambda$ one has for the extremal values of the
einbeins:
\be
\mu_1(\lambda)=\sqrt{m_1^2+\Delta^2(\lambda)}\quad\mu_2(\lambda)=\sqrt{m_2^2+\Delta^2(\lambda)}
\quad\mu(\lambda)=\frac12\sqrt{\sigma}\left(\frac{\lambda}{\kappa}\right)^{3/2},
\label{13}
\ee
where 
$$
\Delta^2(\lambda)=\frac{\sigma\lambda}{3\kappa}\left(a+2\lambda\left|\frac{\partial
a}{\partial\lambda}\right|\right).
$$

Then the definition of the reduced field $\mu$ gives the equation for $\lambda$
\be
\mu(\lambda)=\frac{\mu_1(\lambda)\mu_2(\lambda)}{\mu_1(\lambda)+\mu_2(\lambda)},
\label{14}
\ee
which is the subject to numerical investigation. Unfortunately there is no
analytic solution for $a(\lambda)$, so one has to generate a selfconsistent solution
of both equations, (\ref{12}) and (\ref{14}). 

Let us give here two more formulae which will be needed for calculating the
spin-spin and spin-orbit splittings. For radially excited states one can 
find \cite{yuaDB,Lisbon}
\be
\left|\psi(0)\right|^2=\frac{2\mu\sigma}{4\pi}\left(1+\lambda\langle
x^{-2}\rangle\right),
\label{15}
\ee
where
\be
\langle r^N\rangle=(2\mu\sigma)^{N/3}\langle x^N\rangle =
(2\mu\sigma)^{N/3}\int_0^{\infty}x^{N+2}
\left|\chi_{\lambda}(x)\right|^2,\quad N>-3-2l.
\ee

In Table 1 we give the results of numerical calculations with the Hamiltonian
$H_0$ for $D$, $D_s$, $B$ and $B_s$ mesons. 
\begin{table}[t]
\begin{center}
\begin{tabular}{|c|c|c|c|c|c|c|c|c|c|c|c|c|}
\hline
$n$&$l$&meson&$m_1$&$m_2$&$\sigma$&$\alpha_s$&$\lambda$&$\mu_1$&$\mu_2$&$\mu$&$E_0$&$|\psi(0)|$\\
\hline
0&0&$D$&1.4&0.009&0.17&0.4&0.817&1.497&0.529&0.391&2.198&0.161\\
\cline{3-13}
&&$D_s$&1.4&0.17&0.17&0.4&0.847&1.501&0.569&0.412&2.224&0.167\\
\cline{3-13}
&&$B$&4.8&0.005&0.17&0.39&0.999&4.840&0.619&0.549&5.527&0.209\\
\cline{3-13}
&&$B_s$&4.8&0.17&0.17&0.39&1.035&4.842&0.658&0.579&5.550&0.219\\
\hline
0&1&$D$&1.4&0.009&0.17&0.4&0.869&1.522&0.597&0.428&2.640&0\\
\cline{3-13}
&&$D_s$&1.4&0.17&0.17&0.4&0.891&1.525&0.629&0.445&2.663&0\\
\cline{3-13}
&&$B$&4.8&0.005&0.17&0.39&1.052&4.847&0.675&0.593&5.949&0\\
\cline{3-13}
&&$B_s$&4.8&0.17&0.17&0.39&1.080&4.849&0.707&0.617&5.970&0\\
\hline
0&2&$D$&1.4&0.009&0.17&0.4&0.924&1.554&0.674&0.470&2.961&0\\
\cline{3-13}
&&$D_s$&1.4&0.17&0.17&0.4&0.942&1.557&0.702&0.484&2.982&0\\
\cline{3-13}
&&$B$&4.8&0.005&0.17&0.39&1.128&4.860&0.762&0.659&6.245&0\\
\cline{3-13}
&&$B_s$&4.8&0.17&0.17&0.39&1.151&4.861&0.789&0.679&6.263&0\\
\hline
1&0&$D$&1.4&0.009&0.17&0.4&0.929&1.557&0.682&0.474&2.848&0.162\\
\cline{3-13}
&&$D_s$&1.4&0.17&0.17&0.4&0.947&1.561&0.710&0.488&2.869&0.165\\
\cline{3-13}
&&$B$&4.8&0.005&0.17&0.39&1.142&4.863&0.779&0.671&6.131&0.207\\
\cline{3-13}
&&$B_s$&4.8&0.17&0.17&0.39&1.165&4.864&0.806&0.692&6.149&0.212\\
\hline
\end{tabular}
\caption{Solutions of the equations (\ref{12})-(\ref{14}) for standard values of the
string tension $\sigma$, the strong coupling constant $\alpha_s$ and the masses of
the quarks. All parameters are given in $GeV$ to the appropriate powers.}
\end{center}
\end{table}

\subsection{Spin-spin and spin-orbit splittings. Comparison with the
experimental and lattice data}

In this subsection we calculate the contribution of the $V_{string}$ and $V_{sd}$
terms. First of all we collect the averaged values of the spin-orbit and
spin-tensor interactions between various $1P$ and $1D$ eigenstates of the Hamiltonian 
$H_0$
\be
\begin{array}{lll}
\langle ^1P_1|\vec{S}_1\vec{L}|^1P_1\rangle=0&
\langle ^1P_1|\vec{S}_2\vec{L}|^1P_1\rangle=0&
\langle ^1P_1|(\vec{S}_1\vec{n})(\vec{S}_2\vec{n})|^1P_1\rangle=-\frac{\ds
1}{\ds 4}\\
{}\\
\langle ^3P_0|\vec{S}_1\vec{L}|^3P_0\rangle=-1&
\langle ^3P_0|\vec{S}_2\vec{L}|^3P_0\rangle=-1&
\langle ^3P_0|(\vec{S}_1\vec{n})(\vec{S}_2\vec{n})|^3P_0\rangle=-\frac{\ds
1}{\ds 4}\\
{}\\
\langle ^3P_1|\vec{S}_1\vec{L}|^3P_1\rangle=-\frac{\ds 1}{\ds 2}&
\langle ^3P_1|\vec{S}_2\vec{L}|^3P_1\rangle=-\frac{\ds 1}{\ds 2}&
\langle ^3P_1|(\vec{S}_1\vec{n})(\vec{S}_2\vec{n})|^3P_1\rangle=\frac{\ds
1}{\ds 4}\\
{}\\
\langle ^3P_2|\vec{S}_1\vec{L}|^3P_2\rangle=\frac{\ds 1}{\ds 2}&
\langle ^3P_2|\vec{S}_2\vec{L}|^3P_2\rangle=\frac{\ds 1}{\ds 2}&
\langle ^3P_2|(\vec{S}_1\vec{n})(\vec{S}_2\vec{n})|^3P_2\rangle=\frac{\ds 1}{\ds 20}
\end{array}
\ee
\be
\begin{array}{lll}
\langle ^1D_2|\vec{S}_1\vec{L}|^1D_2\rangle=0&
\langle ^1D_2|\vec{S}_2\vec{L}|^1D_2\rangle=0&
\langle ^1D_2|(\vec{S}_1\vec{n})(\vec{S}_2\vec{n})|^1D_2\rangle=-\frac{\ds
1}{\ds 4}\\
{}\\
\langle ^3D_1|\vec{S}_1\vec{L}|^3D_1\rangle=-\frac{\ds 3}{\ds 2}&
\langle ^3D_1|\vec{S}_2\vec{L}|^3D_1\rangle=-\frac{\ds 3}{\ds 2}&
\langle ^3D_1|(\vec{S}_1\vec{n})(\vec{S}_2\vec{n})|^3D_1\rangle=-\frac{\ds
1}{\ds 12}\\
{}\\
\langle ^3D_2|\vec{S}_1\vec{L}|^3D_2\rangle=-\frac{\ds 1}{\ds 2}&
\langle ^3D_2|\vec{S}_2\vec{L}|^3D_2\rangle=-\frac{\ds 1}{\ds 2}&
\langle ^3D_2|(\vec{S}_1\vec{n})(\vec{S}_2\vec{n})|^3D_2\rangle=\frac{\ds 1}{\ds 4}\\
{}\\
\langle ^3D_3|\vec{S}_1\vec{L}|^3D_3\rangle=1&
\langle ^3D_3|\vec{S}_2\vec{L}|^3D_3\rangle=1&
\langle ^3D_3|(\vec{S}_1\vec{n})(\vec{S}_2\vec{n})|^3D_3\rangle=\frac{\ds
1}{\ds 28}\\
\end{array}
\ee
as well as the transition matrix elements
\be
\begin{array}{ll}
\langle ^1P_1|\vec{S}_1\vec{L}|^3P_1\rangle=\frac{\ds 1}{\ds \sqrt{2}}&
\langle ^1P_1|\vec{S}_2\vec{L}|^3P_1\rangle=-\frac{\ds 1}{\ds \sqrt{2}}\\
{}\\
\langle ^1D_2|\vec{S}_1\vec{L}|^3D_2\rangle=\sqrt{\frac{\ds 3}{\ds 2}}&
\langle ^1D_2|\vec{S}_2\vec{L}|^3D_2\rangle=-\sqrt{\frac{\ds 3}{\ds 2}},
\end{array}
\ee
which lead to mixing of $|^1P_1\rangle$, $|^3P_1\rangle$,
and $|^1D_2\rangle$, $|^3D_2\rangle$ pairs
so that the physical state is subject to the matrix equations of the following
type:
\be
\left|
\begin{array}{cc}
E_1-E&V_{12}\\V^*_{12}&E_2-E
\end{array}
\right|=0
\ee 

The results of our numerical calculations as well as the experimental and
lattice data are given in Tables 2,3. The only fitting parameter we use here is
the constant $C_0$ which takes the following values:
\be
C_0(D)=212MeV\quad C_0(D_s)=124MeV\quad C_0(B)=203MeV\quad C_0(B_s)=124MeV.
\label{22}
\ee

As clearly seen from (\ref{22}) the constant $C_0$ practically does not depend 
on the heavy quark and is completely defined by the properties of the light one.
\begin{table}[ht]
\begin{center}
$$
\begin{array}{|c|c|c|c|c|c|}
\hline
{\rm Meson}&n^{2S+1}L_J&J^P&M_{exp}&M_{theor}&M_{lat}\\
\hline
\hline
D&1^1S_0&0^-&1869&1876&1884\\
\hline
D^*&1^3S_1&1^-&2010&2022&1994\\
\hline
%&1^3P_0&0^+&-&2280&\\
%\hline
D_1&\low{1^1P_1/^3P_1}&\low{1^+}&\low{2420}&2354&\\
\cline{5-5}
&&&&\underline{2403}&\\
\hline
D_2&1^3P_2&2^+&2460&2432&\\
\hline
&1^3D_3&3^-&&\underline{2654}&\\
\cline{2-3}\cline{5-6}
{D^*}'&\low{1^1D_2/^3D_2}&2^-&\low{2637}&\underline{2663}&\\
\cline{5-5}
&&&&2729&\\
\cline{2-3}\cline{5-6}
&2^3S_1&0^-&&\underline{2664}&\\
%\hline
%&1^3D_1&1^-&&2728&\\
\hline
\hline
D_s&^1S_0&0^-&1968&1990&1984\\
\hline
D_s^*&1^3S_1&1^-&2112&2137&2087\\
\hline
D_{1s}&\low{1^1P_1/^3P_1}&1^+&\low{2536}&2471&\low{2494}\\
\cline{5-5}
&&&&\underline{2516}&\\
\hline
D_{2s}&1^3P_2&2^+&2573&2547&2411\\
\hline
\hline
B&1^1S_0&0^-&5279&5277&5293\\
\hline
B^*&1^3S_1&1^-&5325&5340&5322\\
%\hline
%&1^3P_0&0^+&-&5655&\\
\hline
B_1&\low{1^1P_1/^3P_1}&\low{1^+}&\low{5732}&5685&\\
\cline{5-5}
&&&&\underline{5719}&\\
\hline
B_2&1^3P_2&2^+&5731&5820&\\
\hline
&1^3D_3&3^-&&\underline{5955}&\\
\cline{2-3}\cline{5-6}
{B^*}'&\low{1^1D_2/^3D_2}&2^-&\low{5860}&\underline{5953}&\\
\cline{5-5}
&&&&6018&\\
\cline{2-3}\cline{5-6}
&2^3S_1&0^-&&\underline{5940}&5890\\
\hline
\hline
B_s&1^1S_0&0^-&5369&5377&5383\\
\hline
B^*_s&1^3S_1&1^-&5416&5442&5401\\
\hline
B_{1s}&\low{1^1P_1/^3P_1}&1^+&\low{5853}&5789&\low{5783}\\
\cline{5-5}
&&&&\underline{5819}&\\
\hline
B_{2s}&1^3P_2&2^+&&5834&5848\\
\hline
\end{array}
$$
\caption{Masses of the $D$, $D_s$, $B$ and $B_s$ mesons in $MeV$. Lattice
results are extracted from Figures 26,27 and Tables XXVIII,XXIX of \cite{lattice}. 
Symbols $1^1P_1/^3P_1$ and $1^1D_2/^3D_2$ are used to indicate that the
physical states
are mixtures of the $1^1P_1$ and $1^3P_1$ or $1^1D_2$ and
$1^3D_2$ states correspondingly. Underlined figures give masses of the most
probable candidates for the experimentally observed resonances.}
\end{center}
\end{table}
\begin{table}[ht]
\begin{center}
$$
\begin{array}{|c|c|c|c|c|c|c|c|c|}
\hline
{\rm Splitting}&D_s-D&D_s^*-D^*&D^*-D&D_s^*-D_s&B_s-B&B_s^*-B^*&B^*-B&B_s^*-B_s\\
\hline
{\rm Experim.}&99&102&141&144&90&91&46&47\\
\hline
{\rm Theory}&114&115&146&147&100&102&63&65\\
\hline
{\rm Lattice}&100&92&110&103&90&90&30&29\\
\hline
\end{array}
$$
\caption{Splittings for the $D$, $D_s$, $B$ and $B_s$ mesons in $MeV$. Lattice
results are taken from Tables XXVIII,XXIX of \cite{lattice}.}
\end{center}
\end{table}
\section{Discussion and conclusions}
 
Several comments concerning our choice of parameters are in order here.
From Table 1 it is seen that we use practically the same values of the strong
coupling constant $\alpha_s$ for all states. To justify this action let us note
that in the case of a light particle moving in the field of a heavy one the OGE 
interaction between them
depends not on the total mass of the system, but rather on its
size. Then the slight variation of the strong coupling is due to
small differences in the mesons sizes. The exact value of $\alpha_s$ is
chosen to be near its frozen value \cite{frozen} that emphasizes the nonperturbative
nature of the processes responsible for the formation of the heavy-light 
mesons spectrum. The fits are very weakly sensible to variations of the heavy 
quarks mass which lead mainly to rescaling of the overall shift constant $C_0$,
so that the latter remains practically the only fitting parameter as stated
in the preceding section. 

In conclusion we would like to comment on the possible identification of the
resonance $D(2637)$ recently claimed by DELPHI Collaboration \cite{DELPHI}. It was reported to
be extremely narrow, about $15MeV$, that leads to a confusion as such a small
width is inconsistent with the identification of this state as the first radial
excitation ${D^*}'(J^P=1^-)$. It was found in a number of papers 
\cite{discussion} that in spite of the fact that the mass of this state perfectly
coincides with the predictions of quark models, all estimates of its width fail
to give such a small value. In the meantime it was found that widths of 
orbitally excited $D$ mesons with quantum 
\clearpage\noindent
numbers $2^-$, $3^-$ could be
consistent with the reported value \cite{discussion}, but then
the following two points were put forward
as main objections: i) a neighboring slightly more massive state should be
observed as well, ii) quark models predict orbitally excited mesons to be 
at least $50MeV$ heavier than needed. Here we would like to comment on the second
argument. As one can find from the fit given in Table 2, in our model the 
first radial and the second orbital excitations share the same region of 
masses. It is not surprise as the negative mass shift of about $50MeV$ for the
orbitally excited state, missing in standard quark models, is readily delivered
by the string correction given by equation (\ref{10}). This contribution does
not affect radially excited states but is significant for orbital excitations.
It was demonstrated in \cite{MNS} that the account of the proper dynamics
of the QCD string in mesons brings the slope of Regge trajectories to their 
correct values. Appearance of the term (\ref{10}) in the meson Hamiltonian and
its important contribution into the masses of orbitally excited states is
yet another reflection of the general situation that the proper QCD string 
dynamics is extremely important in description of hadronic properties and thus it 
should be taken into account. So we conclude that the mysterious $D(2637)$ 
state indeed can be identified with the orbital excitation $2^-$ or $3^-$ 
rather than with the radial
one $0^-$, that resolves the second objective above. 
This statement may hold true for the corresponding states in the $B$ meson spectrum,
where DELPHI Collaboration also claims a similar state \cite{DELPHI2}.

As far as the first one is
concerned, our model also predicts another orbitally excited state with the
mass $2728MeV$, $i.e.$ $65MeV$ higher than $D_2'$ given in the table, and
its experimental grounds are really not clear.

Our predictions for $D(2654)$, $D(2663)$ and $D(2664)$ lie somewhat higher than 
the experimentally observed value. We find it to be a reflection of a general
lack of the \lq\lq $\mu$-technique" used in this paper, which gives larger errors for
higher excited states. The difference of about $15-20MeV$ between the 
theoretical predictions and the experimental datum $2637\pm 6MeV$ can be 
explained in this way. Development of a systematical approach to the einbein 
fields as variational parameters could shed light on the sources of systematical
errors and possibly to improve the results presented in this paper.
\medskip

The authors would like to thank Yu.A.Simonov for constant interest to the work and
useful discussions, R.N.Faustov for reading the manuscript and for valuable comments, 
and B.L.G.Bakker for providing the numerical
solutions for the reduced Schr{\" o}dinger equation.
Financial support of RFFI grants 00-02-17836 and 00-15-96786 and
INTAS-RFFI grant IR-97-232 is gratefully acknowledged.
  
\end{document}